\documentclass[12pt,preprint]{aastex}

\begin{document}

\title{Broad iron-K$\alpha$ emission lines as a diagnostic of black hole spin}
\author{Christopher~S.~Reynolds\altaffilmark{1} and Andrew~C.~Fabian\altaffilmark{2}}
\altaffiltext{1}{Dept.of Astronomy, University of Maryland, College
Park, MD~20742} 
\altaffiltext{2}{Institute of Astronomy, Madingley Road, Cambridge, 
CB3~OHA, U.K.} 

\begin{abstract}
We address the ability of broad iron emission lines from black hole
accretion disks to diagnose the spin of the black hole.  Using a
high-resolution 3-dimensional MHD simulation of a geometrically-thin
accretion disk in a Pseudo-Newtonian potential, we show that both the
midplane density and the vertical column density of the accretion flow
drop dramatically over a narrow range of radii close to the innermost
stable circular orbit (ISCO).  We argue that this drop of density is
accompanied by a sharp increase in the ionization parameter of the
X-ray photosphere, and that the resulting imprint of the ISCO on the
X-ray reflection spectrum can be used to constrain spin.  Motivated by
this simulation, we construct a simplified toy-model of the accretion
flow within the ISCO of a Kerr black hole, and use this model to
estimate the systematic error on inferred black hole spin that may
result from slight bleeding of the iron line emission to the region
inside of the ISCO.  We find that these systematic errors can be
significant for slowly spinning black holes but become appreciably
smaller as one considers more rapidly rotating black holes.
\end{abstract}



\section{Introduction}

Recent years have brought an increasing realization of the
astrophysical importance of black hole spin.  Ever since the seminal
work of Penrose (1969) and Blandford \& Znajek (1977), it has been
recognized that black hole spin may be an important source of energy,
especially for the powerful relativistic jets seen emerging from many
black hole systems.  However, the importance of black hole spin goes
beyond its role as a possible power source.  The spins of stellar-mass
black holes found in accreting Galactic Black Hole Binaries (GBHBs)
are expected to be natal (King \& Kolb 1999) and give us a window into
the black hole forming core-collapse supernova.  The spin distribution
of the supermassive black hole (SMBH) population (and its dependence
on SMBH mass) encodes the black hole growth history, e.g., the role of
accretion versus mergers (Moderski \& Sikora 1996; Volonteri et
al. 2005).  Thus, across the mass scales, black hole spin gives us a
fossil record of how black holes form and grow.

Of course, we must have robust observational probes of black hole spin
if we are to truly test spin-powered jet models or actually use spin
to probe the formation and growth of black holes.  The most direct
method imaginable is the characterization of the gravitational
radiation emitted as another compact object spirals into the black
hole.  However, in the current (pre- gravitational wave observatory)
era, we must search for signs of black hole spin in the
electromagnetic radiation emitted by the surrounding accretion disk.
The natural waveband to use is the X-ray band given the fact that the
observed X-rays are thought to be predominately emitted from the inner
regions of the accretion flow where the relativistic effects related
to black hole spin are strongest.

The most important effect of black hole spin on the accretion disk
arises from the spin dependence of the location of the innermost
stable circular orbit (ISCO); in Boyer-Lindquist coordinates, $r_{\rm
isco}$ shrinks from $6GM/c^2$ for a non-rotating black hole down to
$GM/c^2$ for a hole that is maximally rotating in the same sense as
the accretion disk.  Thus, for a given black hole mass, the
characteristic temporal frequency, energy release (and hence
temperature), and gravitational redshift of the inner accretion flow
all increase as one considers progressively higher black hole spin.

These considerations lead to the three most widely discussed
techniques for using X-ray data to determine black hole spin.  On the
timing front, the frequency stability of the high-frequency
quasi-periodic oscillations (HFQPOs) seen in the soft intermediate
state of GBHBs strongly suggest a connection to the gravitational
potential and black hole spin (Strohmayer 2001; also see McClintock \&
Remillard [2003] for a general review of HFQPO phenomenology).  While
HFQPOs may eventually provide a robust way of measuring spins in
GBHBs, the lack of a compelling theoretical framework with which to
interpret the frequencies prevents any robust conclusions from being
drawn at the present time.  For example, different and contradictory
constraints on the mass and spin of the black hole result from
assuming that the HFQPOs are manifestations of trapped global g-modes
(Nowak \& Wagoner 1992; Nowak et al. 1997), parametric resonances
(Abramowicz et al. 2003), or Lens-Thirring precession (Merloni et
al. 1999).

The second technique involves modeling the thermal continuum spectrum
from the disk.  Provided that one selects systems that have
optically-thick, geometrically-thin inner accretion disks,
observations can be compared with model thermal continuum spectra
(which include vertical radiation transfer in the disk atmosphere as
well as Doppler and gravitational red/blue-shifts; Davis, Done \&
Blaes 2006).  Spin results from this technique have been reported by
McClintock et al. (2006) and Shafee et al. (2006), although the
contrary results of Middleton et al. (2006) demonstrate the current
fragility of this technique to modeling the non-thermal emission,
particularly when applying it to data from the Proportional Counter
Array (PCA) on the {\it Rossi X-ray Timing Explorer (RXTE)} in which
one only sees the Wien tail of the thermal disk spectrum.  While this
is a powerful method for determining the spin of accreting
stellar-mass black holes (especially when applied to broad-band X-ray
data that extends into the soft X-ray band), the fact that one needs
both an estimate of the black hole mass and a high quality measurement
of the thermal continuum shape makes it difficult to apply to the SMBH
in active galactic nuclei (AGN).  The thermal continuum of AGN disks
is in the UV/EUV region of the spectrum, making a determination of its
shape and normalization extremely susceptible to any errors in the
(large) correction that must be done for dust extinction and
photoelectric absorption.

The third technique for constraining spin uses relativistic broadening
and gravitational redshifting of the fluorescent iron emission line
seen in many GBHBs and AGN (Fabian et al, 1989; Laor 1991; Fabian et
al. 2000; Reynolds \& Nowak 2003; Fabian \& Miniutti 2005).  As one
considers progressively more rapidly rotating black holes, the primary
X-ray emission and hence the iron line emission will be dominated by
regions with higher gravitational redshift, leading to broader and
more highly skewed iron emission lines.  A major advantage of this
technique is that the iron line profiles are completely independent of
black hole mass and hence one can apply this to an AGN in which the
mass is often extremely uncertain.  Although one actually proceeds via
formal fitting of relativistically smeared theoretical disk reflection
spectra to data, the black hole spin constraint essentially results
from a characterization of the maximum redshift of the iron line.
Results using this technique on the Seyfert galaxy MCG--6-30-15 have
been reported by Brenneman \& Reynolds (2006) who report rather
extreme constraints on the dimensionless black hole spin of $a>0.987$
(also see Dabrowski et al. 1997).

Both the thermal continuum and the iron line techniques for
determining black hole spin rely on the assumption that the ISCO
defines an inner edge for the thermally or line emitting region of the
disk (see Krolik \& Hawley 2002 for a general discussion of the
various ``inner edges'' that one can define for an accretion disk).
The robustness of the thermal continuum technique has been explicitly
addressed by Shafee, Narayan \& McClintock (2007) using a global
$\alpha$-viscosity model that explicitly tracks the flow properties
within the ISCO.  Although MHD effects close to the ISCO may
invalidate the assumptions of such $\alpha$-models (e.g., Reynolds \&
Armitage 2001), the results of Shafee et al. (2007) do indeed suggest
that modeling of the thermal continuum can produce an accurate measure
of black hole spin.

The subject of the current {\it Paper} is to address the robustness of
the iron line fitting technique.  In Section 2, we present a high
resolution 3-dimensional magnetohydrodynamic (MHD) simulation of a
geometrically-thin ($h/r\sim 0.05$) accretion disk that we use to
explore the nature of the flow close to the ISCO.  In Section 3, we
show that both the midplane density and vertically-integrated surface
density of the accretion flow drops precipitously over a small range
of radii close to the ISCO.  We then argue in Section 4 that
photoionization of the flow within the ISCO will effectively suppress
the iron line emission (and all other X-ray reflection signatures)
across most of the region within the ISCO.  We discuss the
implications and limitations of our results in Section 5, including an
estimate of the uncertainties in the inferred black hole spin that
might arise from iron line emission in the region immediately within
the ISCO.  Section 6 presents our conclusions.

\section{An MHD simulation of a thin accretion disk}

Relativistically broad iron lines are only seen from accretion flows
that are relatively cold (i.e., there are recombined iron ions in the
plasma) and Compton-thick.  Thus, we expect systems displaying broad
iron lines to have radiatively-efficient, geometrically-thin accretion
disks of the type envisaged by Pringle \& Rees (1972), Shakura \&
Sunyaev (1973) and Nokivov \& Thorne (1974).

Our goal is to simulate such an accretion flow in order to assess the
nature of the flow around the ISCO.  Of course, such black hole
accretion disks are very complex systems.  It is believed that the
underlying angular momentum transport that actually allows accretion
to proceed is due to sustained MHD turbulence driven by the
magneto-rotational instability (MRI; Balbus \& Hawley 1991, 1998).
Magnetic energy is dissipated and thermalized through magnetic
reconnection on small scales within the body of the disk.  In the
systems under consideration here, the accretion disk is
optically-thick and the resulting thermal radiation comes to dominate
the gas pressure and, indeed, provides the principal vertical pressure
support for the disk.  The radiation diffuses vertically out of the
disk and is radiated from the disk atmosphere.  Finally, this physics
is occurring in a region where both special and general relativistic
effects are strong and can have qualitative effects.  The most obvious
general relativistic effect is the existence of the ISCO.  However,
frame-dragging, light bending, and the strong displacement currents
that accompany the relativistic fluid velocities can all have
important effects on disk dynamics (e.g., see fully relativistic MHD
simulations of De~Villiers et al. 2004, McKinney \& Gammie 2004).

Global accretion disk simulations that include all of these physical
processes are still beyond current computational capabilities.
Moreover, since we are motivated to model geometrically-thin accretion
disks, the need to spatially resolve the vertical structure of the
disk while still modeling a significant range of radii imposes severe
resolution requirements as will be discussed in more detail below.
Clearly, compromises must be made.

These considerations lead us to construct a model of a thin accretion
disk employing the simplest possible physics that will enable us to
study the structure of the flow around the ISCO.  We employ
3-dimensional MHD in order to correctly capture angular momentum
transport within the flow due to MRI-driven turbulence.  Given that
fully relativistic simulation at the required resolution would prove
extremely computational intensive, together with the current lack of
an adequate publicly available 3-d GRMHD code, we use the
Pseudo-Newtonian potential of Pacynski \& Witta (1980),
\begin{equation}
\Phi=\frac{GM}{r-2r_g},
\end{equation}
where $r_g=GM/c^2$ is the gravitational radius of the black hole which
has mass $M$.  This potential has an ISCO at $r_{\rm isco}=6r_g$ and,
in fact, is a good approximation for the gravitational field of a
non-rotating black hole.  Although we essentially integrate the
equations of non-relativistic MHD within this potential, we use the
prescription of Miller \& Stone (2000) to include some effects of the
displacement current, principally forcing the Alfv\'en speed to
correctly limit to the speed of light as the magnetic fields becomes
strong.  We note that this ``Alfv\'en speed limiter'' only plays a
role within the tenuous magnetized atmosphere that forms at large
vertical distances above and below the disk; it never plays a
significant role in the body of the accretion disk.  Furthermore, gas
velocities within our simulation are always very subluminal (achieving
a maximum of $\sim 0.2c$ at the inner boundary of the simulation
domain).  Thus, we believe that the neglect of the full set of
relativistic terms in the MHD equations should not significantly
impact our simulation.  We also note that De~Villiers \& Hawley (2003)
explicitly comment upon the similarity of pseudo-Newtonian simulations
with full GRMHD simulations performed within a Schwarzschild metric.

In another major simplification, we neglect radiation processes
(radiative heating, radiative cooling and radiation pressure).  In
place of a full energy equation, the gas is given an adiabatic
equation of state with $\gamma=5/3$.  Our simulation does not,
however, capture the magnetic energy lost to numerical
(i.e. grid-scale) reconnection.  While not particularly physical, this
does acts as an effective cooling term for the disk and helps maintain
a thin, cold structure.

The simulation is performed in cylindrical polar coordinates
$(r,z,\phi)$.  The initial condition for the simulation consists of a
disk with a constant mid-plane density $\rho(r,z=0)=\rho_0$ for
$r>r_{\rm isco}$.  The vertical run of initial density and pressure
correspond to an isothermal atmosphere in vertical hydrostatic
equilibrium,
\begin{eqnarray}
\rho(r,z)&=&\rho_0\,\exp\left(-\frac{z^2}{2h(r)^2}\right),\\
p(r,z)&=&\frac{GMh(r)^2}{(R-2r_g)^2R}\,\rho(r,z),
\end{eqnarray}
where $r$ is the cylindrical radius, $z$ is the vertical height above
the disk midplane and $R=\sqrt{r^2+z^2}$.  The scale height of the
disk as a function of radius is a free function; we choose to set
$h(r)=0.3r_g$ for all $r$, i.e., a constant height disk with initial
$h/r=0.05$ at the ISCO.  The initial density is set to zero for
$r<r_{\rm isco}$.  The initial velocity field is everywhere set to
\begin{equation}
v_\phi=\frac{\sqrt{GMr}}{r-2},\hspace{1cm}v_r=v_z=0,
\end{equation}
corresponding to rotation on cylinders and pure Keplerian motion for
material on the mid-plane.  An initially weak magnetic field is
introduced in the form of poloidal field loops specified in terms of
their vector potential ${\bf A}=(A_r,A_z,A_\phi)$ in order to ensure
that the initial field is divergence free.  We choose the explicit
form for the vector potential,
\begin{eqnarray}
A_\phi=A_0\,f(r,z)\,p^{1/2}\,\sin \left(\frac{2\pi r}{5h}\right),\hspace{1cm}A_r=A_z=0,
\end{eqnarray}
where $A_0$ is a normalization constant and $f(r,z)$ is an envelope
function that is unity in the body of the disk and then smoothly goes
to zero at both $r=r_{\rm isco}$ and at a location three pressure
scale heights away from the midplane of the disk.  The use of $f(r,z)$
keeps the initial field configuration well away from either the inner
radial boundary of the initial disk configuration or the vertical
boundaries of the calculation domain.  The final multiplicative term
produces field reversals with a radial wavelength of $5h$.  This
results in a number of distinct poloidal loops throughout the disk and
acts as a much more effective seed for the MRI than would a single
loop which followed contours of pressure (in the sense that a
steady-state is reached more quickly).  The normalization constant
$A_0$ is set to give an initial ratio of gas-to-magnetic pressure of
$\beta=10^3$ in the inner disk.

From the body of literature on MRI driven turbulence (e.g., see Balbus
2003), we expect the final turbulent state to be independent of most
aspects of the initial field configuration (e.g., see the study of
Hawley 2000).  In general terms, the one aspect of the initial field
that can continue to leave imprints on the turbulent state is the net
flux threading the simulation domain (e.g., see Reynolds \& Armitage
2001).  Our initial condition has zero net flux, a choice that is
particularly well defined and robust to implement but is expected to
minimize the saturation field strength and the turbulent angular
momentum transport.  It is beyond the scope of this paper to explore
disks that have a net poloidal field resulting, for example, from
advection of magnetic flux from the mass reserviour.

Apart from the Miller \& Stone (2000) Alfv\'en speed limiter, grid
effects and the usual artificial viscosity, this simulation deviates
from strict ideal MHD in one other way.  We impose a floor to the
density field of $10^{-5}$ times the initial maximum density in order
to prevent the numerical integration from producing negative
densities.  This essentially amounts to a subtle distributed mass
source.  The density only reaches this floor close to the $z$-boundary
(i.e., many scale heights above and below the disk plane).

From these initial conditions, the equations of ideal MHD were evolved
using an MPI parallelized version of the ZEUS code (Stone \& Norman
1992a,b for a discussion of the basic ZEUS algorithm).  The
computational domain was the cylindrical wedge defined by $r\in
[4,16]$, $z\in [-1.5,1.5]$ and $\phi\in [0,\pi/3]$.  The computational
grid was uniformly spaced in $(r,z,\phi)$ with $n_r\times n_z\times
n_\phi=480\times 256\times 64$.  This gridding provides $\sim 25$ grid
cells per vertical pressure scale-height in the body of the disk,
adequate to capture the MRI and the large-scale properties of the MHD
turbulence that it drives, while also giving a radial grid spacing
that is only a factor of two larger than the vertical grid spacing.
It is highly desirable to avoid large aspect ratio computational cells
due to the anisotropic numerical viscosity and resistivity that it
would introduce.

Periodic boundary conditions were imposed on the $\phi$-boundaries,
and ``zero-gradient'' outflow boundary conditions were imposed at both
the inner and outer radial boundaries (Stone \& Norman 1992a,b).  The
choice of the $z$-boundary condition is less clear.  Experiments
employing outflow boundaries in the $z$-direction showed that tenuous
matter generally flows slowly across these boundaries at very
sub-sonic and sub-Alfv\'enic speeds; strictly, this invalidates the
use of such boundary conditions (since the flow on the other side of
the boundary should be able to act back on the simulation domain).  We
also note that the use of this outflow boundary condition together
with the imposition of the density floor led to occasional numerical
instabilities at the $z$-boundaries (in the form of a small and
accelerating magnetized vortex) that, while having no effect on the
body of the accretion disk, would eventually halt the simulation.
Instead, we choose to employ periodic boundary conditions in the
$z$-directions.  While this is obviously unphysical in the sense that
matter cannot leave the simulation domain in the vertical direction,
it does guarantee mathematically reasonable behaviour at the boundary
(eliminating numerical instabilities) and, more importantly, appears
to have no effect on the dynamics of the accretion disk itself (as
diagnosed through comparisons with our vertical-outflow runs).

\begin{figure}[t]
\centerline{
\hspace{-0.5cm}
\includegraphics[width=0.9\textwidth]{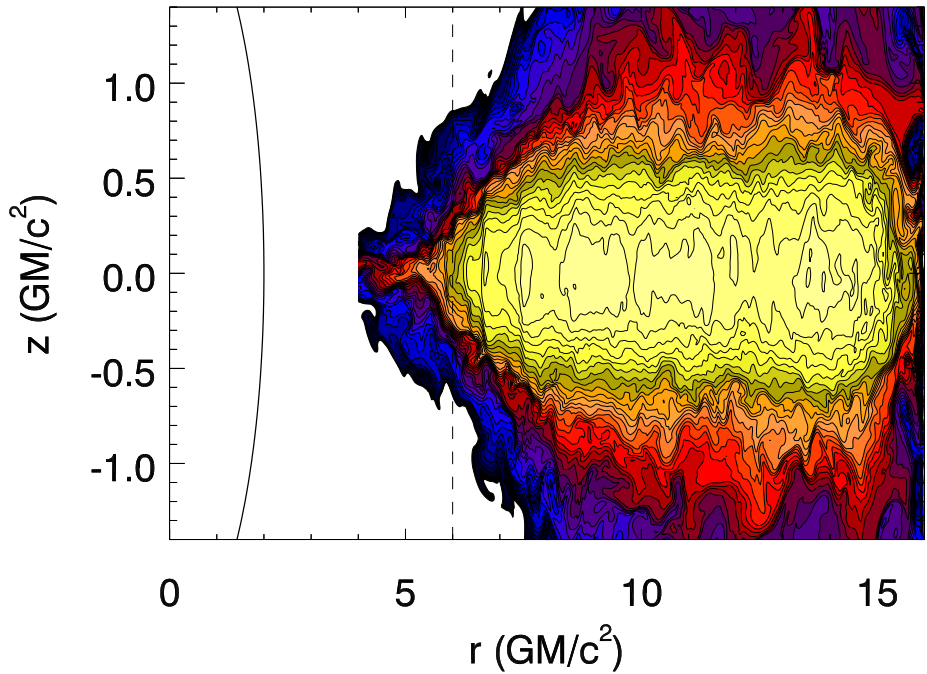}
}
\caption{Logarithmic density across an azimuthal cross-section of the
simulation at time $t=35T_{\rm isco}$.  There are 10 contours per
decade of density.  The distinct ridges at approximately $6.5r_g$ and
$7.5r_g$ are cuts through transient spiral density waves.  The
simulation domain has an inner radial boundary at $r=4r_g$.  However,
the figure shows an extension of this domain to $r=0$ along with a
representation of the event horizon (thick solid black line).  Notice
the rapid transition of the high density disk into a
geometrically-thin, low density accretion stream close to the ISCO.  }
\label{fig:snapshot}
\end{figure}

The initial state undergoes rapid evolution as radial pressure forces
push matter into the (numerical) vacuum within the ISCO.  This drives
axisymmetric, outward-radially directed gravity waves which rapidly
break to become outward traveling rolls.  These hydrodynamic
transients are short lived, lasting for only $5-10T_{\rm isco}$, where
$T_{\rm isco}=61.6GM/c^3$.  Concurrently, MRI amplifies the initial
magnetic field until the volume-averaged $\beta\sim 50$ at which point
the flow becomes turbulent.  The simulated disk is turbulent at all
radii after $t=10T_{\rm isco}$.  Over the subsequent 7 orbits, the
$\beta$ of the disk declines a little due to the effects of magnetic
buoyancy.  Examination of the mass accretion rate and total magnetic
and thermal energy of the disk show that a quasi-steady state
(characterized by approximately constant total magnetic and thermal
energies) is achieved after approximately $t=17T_{\rm isco}$ orbital
periods.  In this steady state, the dense body of the disk (i.e.,
close to the mid-plane) is characterized by $\beta\sim 100$.  While
this is significantly larger than the $\beta$ found in global
simulations of thicker disk, it is in line with what might be expected
for thin accretion disks with zero net field as diagnosed through
local shearing-box simulations (Hawley, Gammie \& Balbus 1996).  We
follow the simulation for a total of 40 ISCO orbital periods in order
to confirm that it is, indeed, in a quasi-steady state.
Figure~\ref{fig:snapshot} shows the density of the disk across an
azimuthal cross-section at a time $t=35T_{\rm isco}$.

\section{Properties of the flow close to the ISCO}

\begin{figure}
\begin{center}
\includegraphics[width=0.8\textwidth]{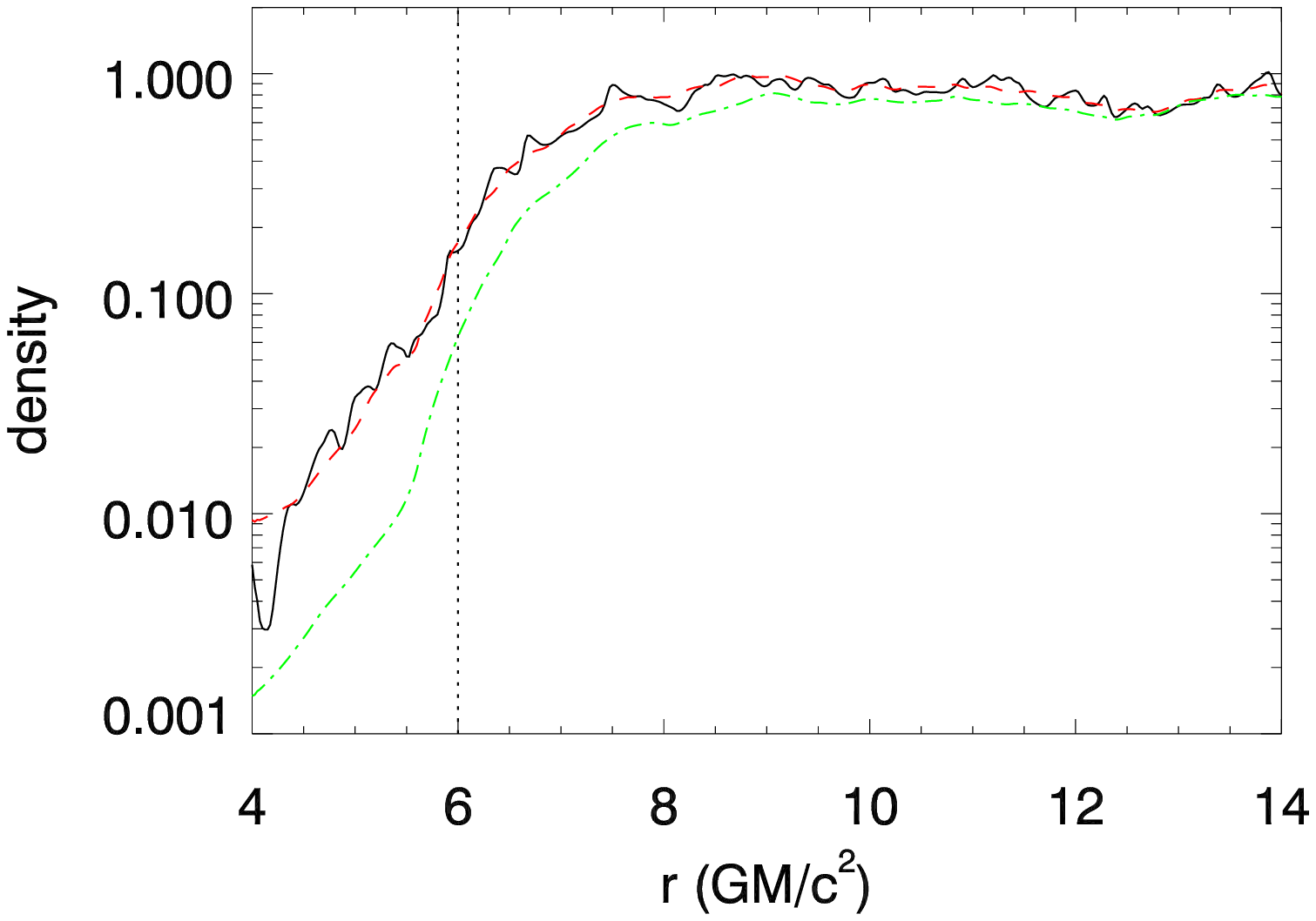}
\includegraphics[width=0.8\textwidth]{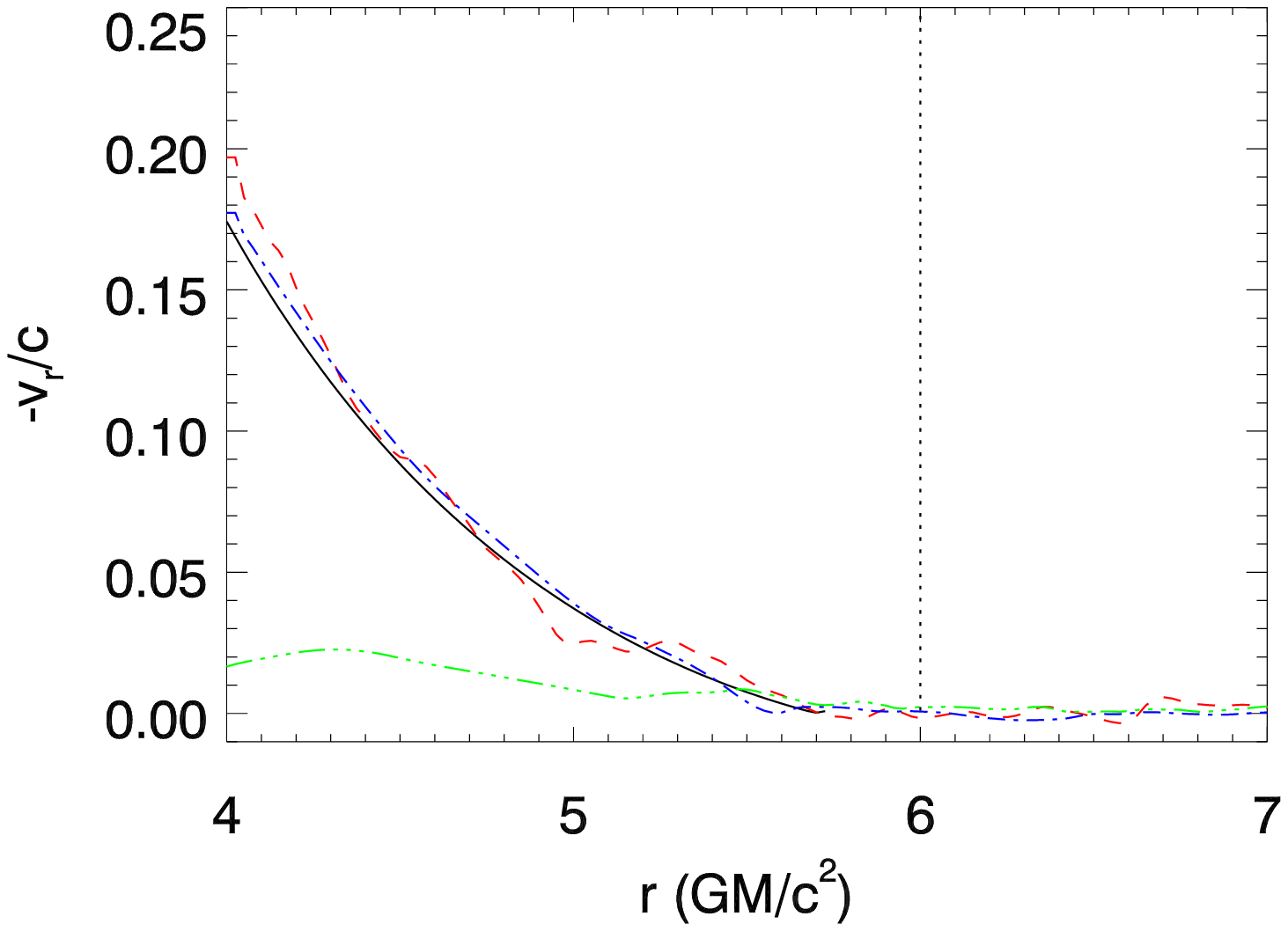}
\end{center}
\caption{{\it Top panel : }Mid-plane density along a single ray (black
solid line), azimuthally-averaged mid-plane density (red dashed line)
and (normalized) total vertical column density (green dot-dashed
line).  The dotted vertical line shows the location of the ISCO. {\it
Bottom panel : }Mid-plane radial velocity along a single ray (red
dashed line), azimuthally-averaged mid-plane radial velocity (blue
dot-dashed line), and the azimuthally-averaged Alfv\'en speed (green
dot-dashed line).  Also shown is the radial velocity of a ballistic
trajectory with the energy and angular momentum corresponding to a
circular orbit at $5.7r_g$ (solid black line).  All quantities are
determined at $t=35T_{\rm isco}$.}
\label{fig:density}
\end{figure}

Using this simulation, we can examine the nature of the flow close to
the ISCO.  Figure \ref{fig:density} (top) shows the mid-plane density
and total vertical column density through the disk as a function of
$r$.  We see that both the midplane density and total column density
undergo a rapid decrease over a small range of radii as the flow
crosses the ISCO; the mid-plane density drops by a factor of 20 and
the column density drops by a factor of 60 as between $r=7r_g$ and
$r=5r_g$.  Just within the ISCO, the radial scale-length for the
mid-plane density and the column density (i.e., the change in radius
over which these quantities drop by a factor $e$) varies from
$0.2-0.5r_g$, i.e., within a factor of two of the vertical scale-height
of the disk outside of the ISCO.  In Section~4, we will show that this
precipitous drop in density close to the ISCO effectively truncates
the iron line emission (i.e., using the terminology of Krolik \&
Hawley [2002], we argue that the disk has a ``reflection edge'' that is
close to the ISCO).

As expected from simple continuity, this density drop accompanies a
rapid increase in the (inward) radial velocity of the gas (bottom
panel of Fig.~\ref{fig:density}).  The azimuthally-averaged midplane
radial velocity is very close to that of a ballistic infall starting
from a circular orbit at $5.7r_g$ (i.e., inset from the ISCO by an
amount comparable to the vertical pressure scale-height of the disk
just beyond the ISCO).  Interestingly, the inflow becomes
super-Alfv\'enic within $r=5.5r_g$, implying that MHD forces can no
longer transport energy and angular momentum upstream within this
radius.  This is the ``stress edge'' of the disk, as defined by Krolik
\& Hawley (2002).  The flow is essentially ballistic within this
radius.  A comparison of this result with those of Hawley \& Krolik
(2001) supports the conjecture (Armitage, Reynolds \& Chiang 2000;
Afshordi \& Paczynski 2003) that the magnetic extraction of energy and
angular momentum from within the ISCO (Krolik 1999; Gammie 1999)
becomes less important as one considers progressively thinner
accretion disks, at least for the initially zero net-field case.

\begin{figure}[t]
\centerline{
\includegraphics[width=0.8\textwidth]{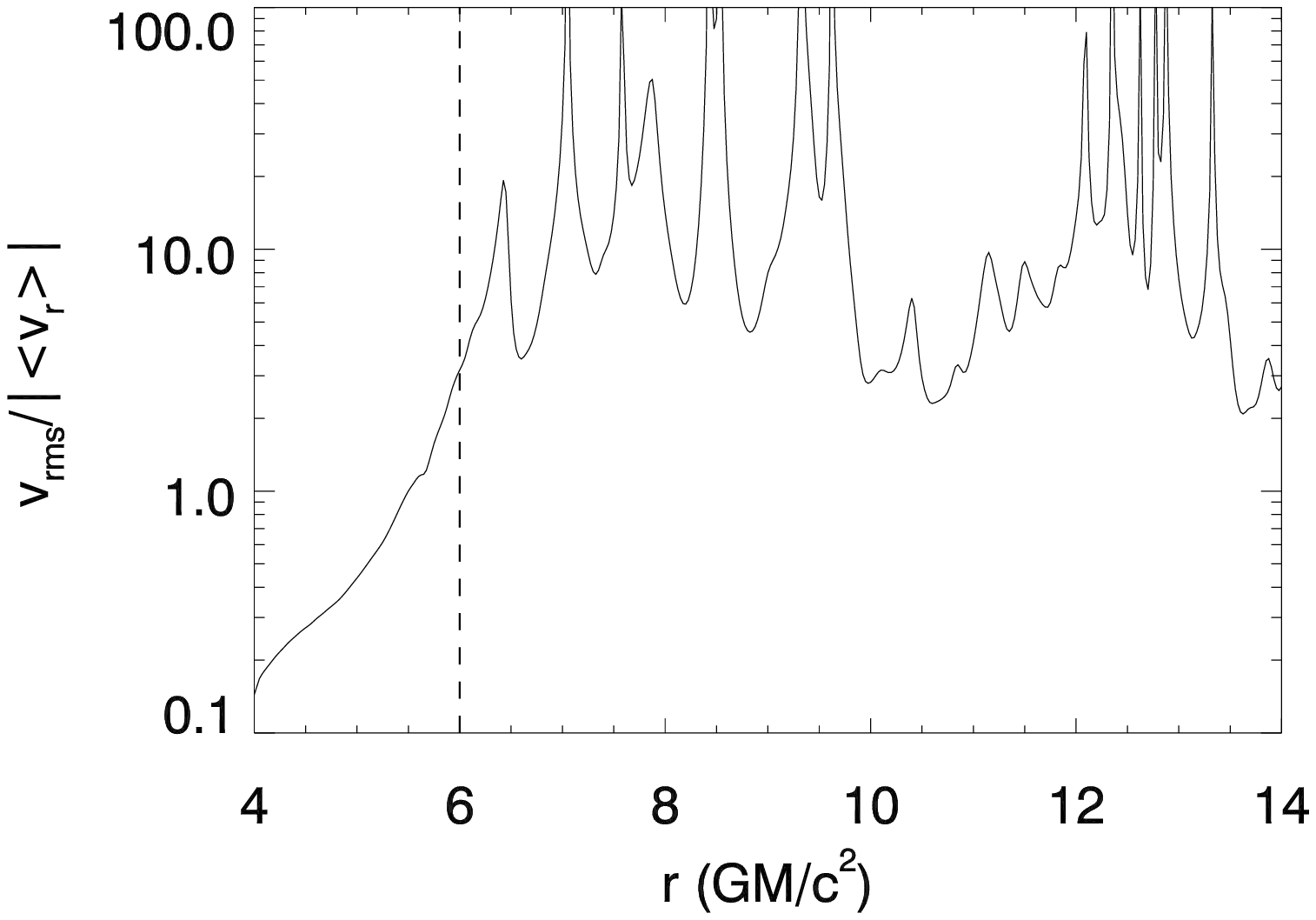}
}
\caption{Ratio of the density-weighted rms fluctuations of the radial
velocity about the (density-weighted) average radial velocity against
the average radial velocity.  The disk undergoes a transition from a
turbulence dominated regime (high values of this ratio) to a more
laminar inflow regime (low values of this ratio).  The formal
``turbulence edge'' where this ratio is unity occurs at $r=5.52r_g$. }
\label{fig:turb_edge}
\end{figure}

Another radius of interest is the ``turbulence edge'', i.e., the
radius at which the flow transitions from being a turbulence dominated
flow to a more laminar radial inflow.  Following Krolik \& Hawley
(2002), we examine the turbulence edge by comparing the
density-weighted azimuthally and vertically-averaged radial velocity
$\bar{v_r}$ to the density-weighted root mean square fluctuations of
$v_r$ around that average (see Fig.~\ref{fig:turb_edge}).  This figure
reveals that the disk does not start feeling the effect of the
plunging region until $r=7r_g$, and that the formal turbulence edge at
$r\approx 5.5r_g$.  Again, we see an abrupt transition close to the
ISCO.

We note that the results discussed in this Section appear to be mostly
insensitive to the numerical resolution and placement of the vertical
boundaries.  Re-running the simulation with half the radial and
vertical resolution leads to 20--30\% increases in the density field
and column density within the ISCO, with a comparable decrease in the
inflow velocity.  A third simulation in which the vertical size of the
computational domain is doubled to $z\in [-3,3]$ (at the reduced
resolution) shows essentially identical density and velocity fields
both within the disk and inside the ISCO.

\section{The use of iron lines as a spin diagnostic}

As discussed in the introduction, the current methodology for using
broad iron lines to constrain black hole spin relies on the assumption
that the iron line emission essentially truncates at the ISCO.  Here,
we employ our MHD simulation to examine this assumption.

\subsection{Physical requirements for iron line production from within the 
ISCO}

The physical requirements for producing significant iron line emission
from within the ISCO were first addressed by Reynolds \& Begelman
(1997; hereafter RB97) using a simple one-zone ballistic model for the
accretion flow.  There are three principal considerations.  Firstly,
significant iron line production requires the flow to be Thomson-thick
in the vertical direction.  For a steadily accreting source with a
radiative efficiency $\eta$, conservation of baryon number gives the
electron scattering optical depth of the plunging region to be,
\begin{equation}
\tau_{\rm e}=\frac{2cr_g}{\eta r (-u^r)}{\cal L},
\label{eq:tau}
\end{equation}
where $u^r$ is the radial component of the 4-velocity, and ${\cal
L}=L/L_{\rm Edd}$ is the Eddington ratio of the source.  Evaluating
this at the event horizon of a Schwarzschild black hole (where
$u^r\approx -c$ and $r_{\rm h}=2r_g$), we see that the entire flow
within the ISCO is Thomson-thick provided that $L>\eta L_{\rm Edd}$.
Using the canonical Schwarzschild value of $\eta\approx 0.06$, we see
that the entire flow is optically-thick provided that it is accreting
at 6\% or more of its Eddington limit, a condition that is very likely
satisfied for the active AGN and high-state GBHBs in which we find
relativistic broad iron lines.  Of course, for more rapidly rotating
black holes, the radiative efficiency is higher (achieving a
theoretical maximum of $\eta=0.42$ for a black hole with $a=1$)
leading to the conclusion that the accretion flows around rapidly
spinning black holes may have an inner optically-thin region unless
the Eddington ratio is close to unity.

The second requirement for significant iron line emission within the
ISCO is that this region is irradiated by a hard X-ray source.  RB97
adopted the optimal geometry, a hard X-ray source that is located
above the black hole, on or close to the symmetry axis of the system.
Observational support for just such a geometry has emerged from
detailed studies of iron line and reflection bump variability by {\it
XMM-Newton} (Fabian \& Vaughan 2003; Miniutti \& Fabian 2004) and {\it
Suzaku} (Miniutti et al. 2007).  

The third and most constraining requirement is that, despite the
irradiation by the hard X-ray source (which may be significantly
enhanced by gravitational light-bending) and the plummeting density,
the surface layers of the plunging flow must maintain a sufficiently
low ionization state to still produce iron features.  RB97 found this
to be possible only in the case where a very small fraction of the
accretion power ($10^{-3}\dot{M}c^2$ or less) is channeled into the
non-thermal X-ray source.  Even in this case, the emission line from
within the ISCO is mostly from H-like and He-like iron.  Young, Ross
\& Fabian (1998) showed that a significant contribution of such highly
ionized iron line emission from within the ISCO would be accompanied
by a deep K-shell absorption edge of ionized iron that is absent from
the data.  Of course, even a fully ionized plunge region will still
reflect incident X-ray photons --- in this case, the interaction is
dominated by Compton scattering and, for X-ray energies under
consideration here, the reflected spectrum has the same shape as the
incident continuum.  When interpreting time-averaged spectra, one
could not distinguish between direct primary X-ray emission and
primary X-ray emission that has been reflected off the fully ionized
flow.  However, spectral variability and time-lags might reveal such a
component.

\subsection{Simulation results}

We now reassess the ionization state of the accretion flow within the
ISCO, and its ability to produce significant amounts of iron line
emission, in the light of the high-resolution MHD simulation presented
in Sections~2 and 3.  As a preliminary to this discussion, we must
dimensionalize our simulation.  For an assumed Eddington ratio ${\cal
L}$, we can apply eqn.~\ref{eq:tau} to the flow at the inner radial
edge of the simulation to relate the (dimensionless) column density in
the simulation to the physical column density in the corresponding
``real'' accretion flow.  Further, once the mass $M$ of the central
object is chosen, the length and velocity scale of simulation is set,
allowing one to dimensionalize the density and pressure.

The X-ray reflection spectrum is sensitive to the ionization state of
the matter at the X-ray ``photosphere'' of the accretion disk,
situated approximately at the $\tau_{\rm e}=1$ surface.  For the
conditions relevant to these irradiated disk atmospheres,
photoionization dominates collisional ionization.  Furthermore, the
photoionization and recombination timescale of the plasma in both the
disk and the plunge region is always very short.  Hence, the
ionization state of the plasma can be described by an ionization
parameter,
\begin{equation}
\xi\equiv \frac{4\pi F_{\rm i}}{n_{\rm ph}},
\end{equation}
where $F_{\rm i}$ is the ionizing ($E>13.6$\,eV) radiation flux and
$n_{\rm ph}$ is the electron number density of the plasma at the
photosphere.  

The density and hence ionization state of the plasma at the
photosphere will very much depend on the vertical structure of the
accretion disk.  While we expect our simulation to capture well the
radial angular momentum transport and hence the radial structure of
the disk around the ISCO, we are forced to neglect physics that is
important for determining the vertical structure.  Most importantly,
we fail to capture the radiative cooling of the disk material, as well
as the role of radiation pressure and radiative transport in the
vertical structure.  These radiation effects are just recently being
included in local (shearing box) simulations of accretion disks (e.g.,
see Turner 2004; Blaes, Hirose and Krolik 2007) and it is not yet
feasible to perform the corresponding global disk simulations.  Thus,
we must attempt to diagnose the ionization state of the disk
atmosphere despite the knowledge that the vertical structure of the
simulated atmosphere is incorrect.  

We parameterize this uncertainty via the ratio of the density at the
photosphere to the average density, $g(r)$, defined by
\begin{equation}
n_{\rm ph}\equiv \frac{\tau_e}{2h\sigma_{\rm T}}g,
\end{equation}
where $h$ is the scale height defined as the $z$-value at which the
density falls to $1/e$ of its midplane value.  We shall refer to
$g(r)$ as the {\it photospheric density parameter}.  We expect the
radial dependence of $\tau_e$ and $h$ close to the ISCO to be
determined from the radial dynamics and hence be well described by
this simulation.  We discuss constraints on $g(r)$ below.

In order to examine the radial run of ionization parameter, we must
specify the illuminating ionizing radiation flux as a function of
radius.  For concreteness (and following RB97), we shall consider a
source of ionizing flux situated at a height $D_{\rm s}$ on the
$z$-axis.  In the absence of any relativistic effects, the resulting
illumination pattern (assuming a flat thin disk) would be
\begin{equation}
F_{\rm i, non-rel}=\frac{L_{\rm i}D_{\rm s}}{(r^2+D_s^2)^{3/2}},
\end{equation}
where $L_{\rm i}$ is the ionizing luminosity of the source.  This form
is modified due to gravitational light-bending, and the gravitational
and Doppler redshifting/blueshifting of source photons (including the
$k$-correction that brings photons into or takes them out of the
ionizing regime).  For $D_{\rm s}\sim r_{\rm isco}$, the result is to
remove the flattening of the illuminating profiles --- the resulting
profile typically {\it steepens} from $r^{-3}$ to approximately
$r^{-4}$ as one considers smaller and smaller radii (Fukumura \&
Kazanas 2007).

In what follows, we shall make the conservative assumption that the
illumination law maintains a $r^{-3}$ dependence at all radii
\begin{equation}
F_{\rm i}=\frac{L_{\rm i}D_{\rm s}}{4\pi r^3}.  
\end{equation}
This is conservative in the sense that it under-illuminates the
innermost regions of the disk (including the region within the ISCO)
and hence produces an under-estimate of the ionization parameter in
the plunge region.  Thus, the final expression that we use to estimate
an ionization parameter from our simulation is
\begin{equation}
\xi=\frac{2 L_{\rm i}h D_{\rm s}\sigma_T}{\tau_e r^3}\frac{1}{g},
\label{eq:xi}
\end{equation}
which, when scaled to typical parameters relevant to a luminous
Seyfert galaxy or a high-state GBHB gives,
\begin{eqnarray}
\xi=1.22\times
10^2\,\left(\frac{h/r}{0.1}\right)\left(\frac{D_s}{r}\right)\left(\frac{\tau_e}{10^3}\right)^{-1}\\
\times \left(\frac{r}{10r_g}\right)^{-1}\left(\frac{{\cal
L}_i}{0.01}\right)\frac{1}{g}\ {\rm erg}\,{\rm cm,{\rm
s}^{-1}}\nonumber
\label{eq:scaledxi}
\end{eqnarray}
where ${\cal L}_i=L_i/L_{\rm Edd}$ is the Eddington ratio for just the
ionizing luminosity.  Note that eqn.~\ref{eq:tau} is only strictly
meaningful inside the plunge region (where the directed radial inflow
exceeds the turbulent fluctuations) and hence it would be
inappropriate to use eqn.~\ref{eq:tau} to simplify
eqn.~\ref{eq:scaledxi}.

To demonstrate the problems encountered if we naively adopt the
vertical structure of the simulated disk, let us examine the
ionization in the disk at $r=10r_g$.  Observations tell us that these
radii in the accretion disks of moderately luminous Seyfert galaxies
(with ${\cal L}\sim 0.1$ and ${\cal L}_i\sim 0.02$) are capable of
producing ``cold'' 6.4\,keV iron lines, implying that $\xi<100\ {\rm
erg}\,{\rm cm},{\rm s}^{-1}$.  Employing the $t=35T_{\rm isco}$
snapshot, the simulation gives $h/r\approx 0.047$ and, for this
Eddington ratio, a total optical depth of $\tau_e\approx 2.5\times
10^3$.  However, integrating a ray into the disk, we hit the
$\tau_e=1$ surface at $z/r\approx 0.14$ giving a photospheric density
parameter $g\approx 5.2\times 10^{-3}$.  For $D_s=6r_g$, this gives an
ionization parameter of $\xi({\rm r=10})\approx 5.2\times 10^3\ {\rm
erg}\,{\rm cm}\,{\rm s}^{-1}$.  Such a high ionization parameter would
render the disk completely incapable of producing iron line features,
contrary to observations.  The problem arises due to the low-density,
high-temperature atmosphere of gas that envelopes the simulated
accretion disk. In real disks, we expect radiative cooling
(principally Compton cooling) to collapse this tenuous atmosphere back
towards the dense parts of the disk, thereby increasing the plasma
density at the photosphere and allowing iron line features to be
emitted.  The density at the photosphere would be further increased by
the action of radiation pressure in the optically-thick body of the
disk which tends to flatten the density profile in the optically-thick
body of the disk.  We can see from the numbers above that the
photospheric density parameter must be increased to $g\sim 0.25$ in
order to produce the iron line features found in the observations.

\begin{figure}[t]
\centerline{
\includegraphics[width=0.8\textwidth]{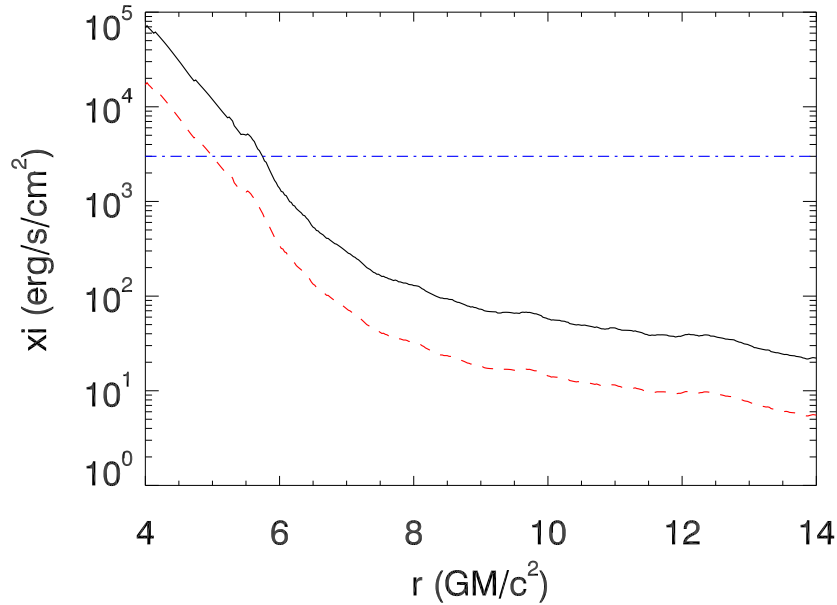}
}
\caption{Run of photospheric ionization parameter as inferred from the
simulation using eqn~\ref{eq:scaledxi}.  For all curves shown,
$D_s=6r_g$ and ${\cal L}_i/{\cal L}=0.1$.  Within this choice of
parameters, these curves are valid for all accretion rates.  Shown
here are the results for $g=0.25$ (black, solid line) and the extreme
case of $g=1.0$ (red, dashed line).  The blue dot-dashed line shows an
ionization state of $\xi=3\times 10^3\ {\rm erg}\,{\rm s}^{-1}\,{\rm
cm}^{-2}$, above which the production of any iron line emission
becomes extremely inefficient.  See main text for further details.}
\label{fig:xi}
\end{figure}

In the absence of global simulations that include this physics, we
must take a semi-empirical approach.  As is clear from above, we need
$g\sim 0.1-1$ outside of the ISCO in order to produce the iron line
features that we see in many Seyfert galaxies and high-state GBHBs.
We take the approach of assuming that $g$ maintains a constant value
of order unity into the plunging region.  Figure~\ref{fig:xi} shows
the run of ionization parameter for $g=0.25$ and the extreme case of
$g=1.0$.  Since $\tau_e\propto {\cal L}$, eqn.~\ref{eq:scaledxi} and
the curves plotted in Fig.~\ref{fig:xi} depend only on the fraction of
the radiation that is ionizing (${\cal L}_i/{\cal L}$) and not the
absolute accretion rate or Eddington ratio.  Note that the
photospheric ionization parameter climbs steeply as one moves inwards
into the plunge region.  The ionized reflection models of Ross \&
Fabian (2005) show that distinguishable iron line features are lost
when $\xi>3\times 10^3 \ {\rm erg}\,{\rm cm}\,{\rm s}^{-1}$.  In fact,
a deep iron absorption edge (which is not present in the data)
accompanies the iron line once $\xi>1\times 10^3 \ {\rm erg}\,{\rm
cm}\,{\rm s}^{-1}$.  Even for the extreme case of $g=1.0$ (i.e., the
photosphere is characterized by the full average density of the disk),
the plunge region becomes incapable of producing relevant iron line
emission within $5r_g$.  For $g=0.25$, the truncation radius is
$5.8r_g$.

\section{Discussion}

In Section~4, we have demonstrated that the presence of the ISCO
leaves a strong imprint on the reflection spectrum of the accretion
disk due to the rapid increase in ionization parameter as the flow
plunges inwards.  The increase in ionization parameter is directly
associated with the decrease in both the midplane density and column
density of the disk as the flow accelerates inwards.  More precisely,
for the particular scenario simulated here, we find that the
transition in the flow properties occurs on a {\it radial} length
scale which is similar to the {\it vertical} scaleheight of the disk.
Our simulation shows that the azimuthally-averaged midplane radial
velocity is very close to that of a ballistic infall starting from a
circular orbit at $5.7r_g$ (i.e., inset from the ISCO by an amount
comparable to the vertical pressure scale-height of the disk beyond
the ISCO).  Indeed, our results explicitly demonstrate that the
turbulence edge, stress edge, and reflection edge all occur within 2
scaleheights of the ISCO.

Thus, the basic principle underlying the use of broad iron lines as a
spin diagnostic is validated.  However, it should also be apparent
that the iron line emission may not truncate {\it precisely} at the
ISCO.  Thus, we must expect to introduce a systematic error in the
black hole spin when we fit observational data with models that do
truncate precisely at the ISCO.  A detailed quantification of these
systematic errors would involve fitting ISCO-truncated iron line
profiles to realistic synthetic data that have been produced from
models that include some contribution from within the ISCO.  Here, we
address this issue using a simpler (if somewhat less constraining)
approach based on the maximum observed redshift of the iron line
features.

Motivated by our simulation, we consider the following (relativistic)
toy-model for the plunge region of an accretion disk around a rotating
black hole.  We consider an accretion disk whose mid-plane lies in the
equatorial plane of a Kerr spacetime with dimensionless spin parameter
$a$.  We suppose that the disk has a vertical scale-height of $h$ just
outside of the ISCO; $h$ and $a$ are considered parameters of this
toy-model.  We assume that the accreting matter follows circular
test-particle orbits within the Kerr spacetime down to a radius
$r=r_{\rm plunge}$.  Motivated by our MHD simulation above, we
suppose that $r=r_{\rm plunge}$ is located a proper radial distance
$h$ inside the ISCO of the Kerr black hole (as measured by an observer
orbiting with the flow at $r=r_{\rm plunge}$).  The flow within
$r=r_{\rm plunge}$ is assumed to follow ballistic trajectories with
the specific energy and specific angular momentum of the circular flow
at $r=r_{\rm plunge}$.  Thus the velocity field of the accretion flow
is defined once we specify $h$ and $a$.  See the Appendix of Reynolds
et al. (1999) for a summary of the relativistic expressions used to
compute this velocity field.

We assume that there is a source of ionizing radiation situated on the
spin axis of the black hole at a height $D_s=r_{\rm isco}$.  Given the
velocity field for the accretion flow described above, we can combine
the conservation of baryon number (eqn.~\ref{eq:tau}) and the relation
between ionization parameter and vertical optical depth
(eqn.~\ref{eq:xi}) in order to deduce that
\begin{equation}
\xi=\frac{4\pi D_s\eta_{\rm
ion}m_pc^2(-u^r)}{gr}\left(\frac{h}{r}\right)_{\rm pl},
\end{equation}
where $\eta_{\rm ion}$ is the efficiency with which the rest mass of
the accretion flow is converted into {\it ionizing} luminosity of the
source.  Then, for assumed values of $D_s$, $\eta_{\rm ion}$, $g$ and
$(h/r)$, we can estimate the radius in the accretion flow becomes too
highly ionized to produce significant iron line emission.  We set this
limiting ionization parameter to be $\xi=3000\,{\rm erg}\,{\rm
cm}\,{\rm s}^{-1}$.

Suppose that one is attempting to constrain the spin of the black hole
through the maximum redshift of the iron line.  If the true spin of
the black hole is $a_{\rm real}$, we use the above toy-model to
determine the innermost radius that contributes to the observable iron
line emission.  Hence, the observed iron line feature will be more
highly redshifted than if it were truncated at the ISCO.  We use the
algorithms and subroutines of Roland Speith (Speith, Riffert \& Ruder
1995) to compute the maximum observed redshift of this iron feature.
We then determine the black hole spin $a_{\rm fit}$ for which the same
maximum redshift would be obtained if the iron line emitting region is
strictly truncated at the ISCO.

\begin{figure}
\includegraphics[width=0.8\textwidth]{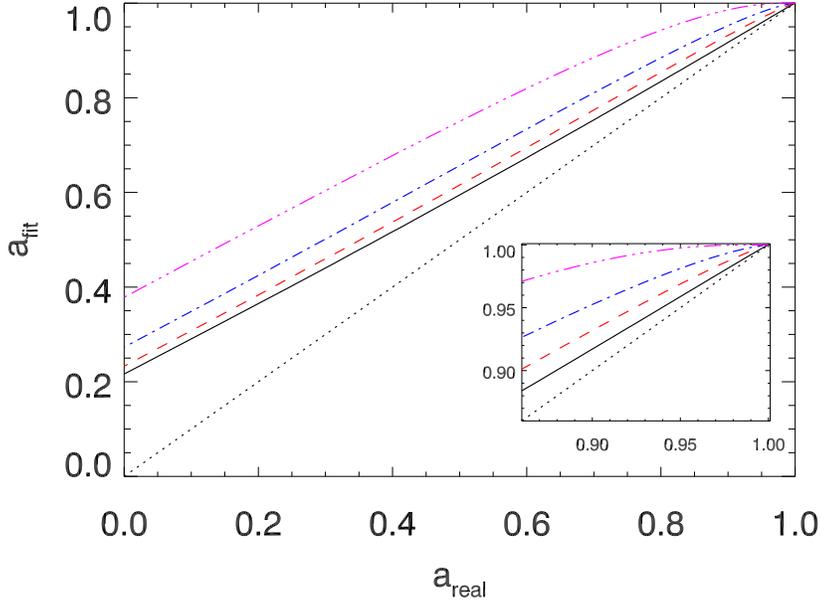}
\includegraphics[width=0.8\textwidth]{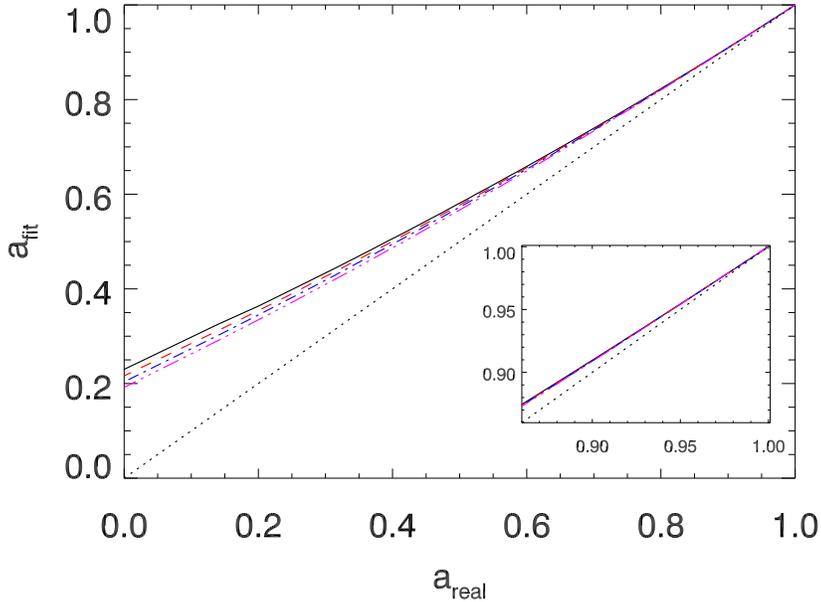}
{\small
\caption{Inferred spin value $a_{\rm fit}$ as a function of actual
spin $a_{\rm real}$ using the maximum redshift method described in the
text; here we assume $g=0.25$, $D_s=r_{\rm isco}$, $(h/r)_{\rm
pl}=0.01$.  We also assume that the accretion flow has the standard
radiative efficiency of a Novikov \& Thorne (1974) accretion disk, and
that 20\% of that luminosity is in the form of ionizing radiation.
The {\it top panel} shows the effect of varying the vertical scale
height of the disk just outside of the ISCO; an inclination of
$i=30^\circ$ is assumed and lines show the very thin case $h=0.01$
(solid black line), $h=0.25r_g$ (red dashed line), $h=0.5r_g$
(dot-dashed blue line), and $h=1.0r_g$ (dot-dot-dot-dashed blue line).
The {\it bottom panel} demonstrates that the viewing inclination of
the accretion disk has only minor impact on these constraints; the
razor-thin disk limit $h=0$ is assumed and lines show the face-on case
$i=0^\circ$ (solid black line), $i=30^\circ$ (red dashed line),
$i=60^\circ$ (dot-dashed blue line), and $i=89^\circ$
(dot-dot-dot-dashed magenta line).}}
\label{fig:spinerror}
\end{figure}

Figure~\ref{fig:spinerror} shows the results of this exercise; the
difference between $a_{\rm fit}$ and $a_{\rm true}$ in these plots
indicates the degree to which the black hole spin can be overestimated
due to the effects of unmodelled iron line emission within the plunge
region.  In generating these figures, we have assumed the canonical
value of the photospheric density parameter $g=0.25$ and a source
which is at $D_s=r_{\rm isco}$.  Motivated by our simulations, we also
assume that the flow within the plunge region has collapsed in the
vertical direction such that $h/r=0.01$.  Finally, we assume that the
total radiative efficiency of the flow is that of a standard Novikov
\& Thorne (1974) accretion disk, and that 20\% of this luminosity is
release in the form of ionizing luminosity from the on-axis source.
This last assumption is probably conservative and motivated by the
multiwaveband observation of the prototypical source MCG--6-30-15
which shows that {\it at least} 20\% of the accretion luminosity is
radiated into the highly-variable X-ray/$\gamma$-ray tail (Reynolds et
al. 1997).

There are several noteworthy aspects of the results reported in
Fig.~\ref{fig:spinerror}.  Firstly, the uncertainty in the black hole
spin parameter introduced by bleeding of the iron line emission region
within the ISCO is generally more important for slowly spinning black
holes.  This is directly related to the fact that both the spacetime
metric and the properties of the accretion disk are slowly varying
functions of black hole spin when $|a|\ll 1$; hence a significant
change in black hole spin is required if an ISCO-truncated iron line
is to mimick the additional redshift introduced by the bleeding of the
line emission region within the ISCO.  

Secondly, the systematic error on the black hole spin increases as one
considers accretion disks that have greater vertical scale height just
outside of the ISCO.  Even for $h\sim r_g$, however, one will still
clearly distinguish rapidly spinning black holes ($a>0.8$) from slowly
spinning holes.  Future global simulations of radiative and
radiation-pressure dominated accretion disks will be required in order
to assess the realistic range of disk thicknesses that characterize
broad iron line sources.  We do note, however, that current
observations impose an important constraint on the geometric thickness
of the inner disk in broad iron line Seyfert galaxies and GBHBs.  For
example, spectral modeling of MCG--6-30-15 shows the inner accretion
disk to have a (photospheric) ionization parameter of $\xi\sim
100\,{\rm erg}\,{\rm cm}\,{\rm s}^{-1}$ (Brenneman \& Reynolds 2006).
Using eqn~\ref{eq:scaledxi} with canonical parameters discussed in
Section 4.2 ($\tau\approx 10^3$, ${\cal L}_i=0.02$, $D_s\approx
6r_g$), we see that this measured ionization parameter along with the
condition $g<1$ demands that $h<0.6r_g$ at $r=6_g$, and $h<0.1r_g$ at
$r=3_g$.  Of course, a more rigorous exploration of this toy-model,
employing relativistically correct treatments of the ionizing
irradiation (beyond the scope of this paper), is required to obtain
reliable limits on the disk thickness for rapidly rotating black
holes.

Thirdly, the precise inclination of the accretion disk has a very weak
effect on the systematic error on the black hole spin (as long as the
inclination is known from, for example, modeling of the overall broad
iron line profile).

Our results suggest that the true spin and the inferred spin become
much closer as one considers more rapidly rotating black holes.  Of
course, for extreme spin values, the inner emitting part of the disk
is very close to the event horizon and well within the ergosphere of
the black hole.  Interactions of the (magnetized) accretion flow with
the strong frame-dragging will likely invalidate the assumptions of
the toy-model presented here.  Significant further work using fully
relativistic codes to explore thin accretion disks is required in
order to assess whether the picture of a well defined transition in
flow properties around the ISCO makes sense for such rapidly rotating
black holes.

\section{Conclusions}

It is a fundamental fact that one can, in principle, observe
arbitrarily large gravitational redshifts from around any black hole,
irrespective of its spin parameter.  Indeed this is very closely tied
to the definition of a black hole.  At first glance, this fact would
appear to render impotent attempts to determine spin using the
gravitational redshift of iron emission lines.  However, it is the
{\it astrophysics} of black holes accretion and the change in the
character of the flow around the ISCO that gives observations of broad
iron lines (and X-ray reflection in general) the power to diagnose
black hole spin.

To understand the behavior of the accretion flow close to the ISCO,
one must capture the angular momentum transport as fully as possible.
Models based on parameterized anomalous viscosity (e.g.,
$\alpha$-models) can become acausal within the plunge region and
cannot capture the non-local MHD stresses that may operate around the
ISCO.  Fully 3-d MHD simulations provide the most powerful tool to
examine these complex flows.  All previously published global 3-d MHD
simulations of black hole accretion disks have revealed rather gradual
transitions of flow properties around the ISCO, but have modeled
rather thick ($h/r\sim 0.1$ or greater) accretion flows.  

Motivated by the fact that luminous Seyfert nuclei and high-state
GBHBs (i.e., the systems in which we actually observe broad iron
lines) are thought to possess geometrically-thin disks, we have
examined the nature of the flow around the ISCO using a
high-resolution pseudo-Newtonian MHD simulation of a disk with
$h/r\sim 0.05$.  We find a rather abrupt transition of flow properties
across the ISCO; the radial length-scale of these changes is
comparable to the vertical scaleheight of the disk beyond the ISCO.
In the language of Krolik \& Hawley (2002), the ``stress edge'' and
the ``turbulence edge'' of the disk are at $r\approx 5.5r_g$ compared
with $r_{\rm isco}=6r_g$.  Most importantly for the present
discussion, we find that both the midplane density and column density
of the disk drop dramatically as the flow accelerates inwards.
Although there is no radiation in the simulation, we relate the
simulation quantities to the expected ionization state of the
accretion flow and find that (even under conservative assumptions
aimed at minimizing the ionization state of the plunging flow)
significant iron line emission cannot be produced from inside
$r=5r_g$.  In fact, for canonical parameters, this ``reflection edge''
is at $r\approx 5.8r_g$, very close to the ISCO.

The fact that the iron line emission does not truncate precisely at
the ISCO will, however, introduce a systematic error into iron line
based spin measurements that assume strictly ISCO-truncated line
profiles.  We make a first attempt to quantify this systematic error
in Section~5, although we note that this approach neglects the spin
information which is carried by the full line profile (as opposed to
just the maximum redshift).  Further study is required to assess
whether, in practice, the information carried by the full line profile
dramatically decreases the systematic error estimates of
Fig.~\ref{fig:spinerror}.

\acknowledgements

We thank Laura Brenneman, Eve Ostriker, Cole Miller, Jon Miller and
John Vernaleo for stimulating discussion related to this project.  We
also thank the anonymous referee for insightful and courteous
comments.  All simulations described in this paper were performed on
the Beowulf cluster (``The Borg'') supported by the Center for Theory
and Computation (CTC) in the Department of Astronomy at the University
of Maryland College Park.  CSR gratefully acknowledges support from
the National Science Foundation under grant AST 06-07428, and the
University of Maryland under the Graduate Research Board Semester
Award Program.  ACF thanks the Royal Society for support.


\begin{references}
\reference{} Abramowicz M.A., Karas V., Kluzniak W., Lee W.H., Rebusco
P., 2003, PASJ, 55, 467

\reference{} Afshordi N., Paczynski B., 2003, ApJ, 592, 354

\reference{} Armitage P.J., Reynolds C.S., Chiang J., 2001, ApJ, 548, 868

\reference{} Balbus S.A., 2003, ARA\&A, 41, 555

\reference{} Balbus S.A., Hawley J.F., 1991, ApJ, 376, 214

\reference{} Balbus S.A., Hawley J.F., 1998, Rev. Mod. Phys., 70, 1

\reference{} Blaes O., Hirose S., Krolik J.H., 2007, ApJ, in press
(astro-ph/0705.0314)

\reference{} Blandford R.D., Znajek R.L., 1977, MNRAS, 179, 433 (BZ)

\reference{} Brenneman L.W., Reynolds C.S., 2006, ApJ, 652, 1028

\reference{} Dabrowski Y., Fabian A.C., Iwasawa K., Lasenby A.N.,
Reynolds C.S., 1997, MNRAS, 288, L11

\reference{} Davis S.W., Done C., Blaes O.M., 2006, ApJ, 647, 525

\reference{} De~Villiers, J.P., Hawley J.F., 2003, ApJ, 589, 458

\reference{} De~Villiers, J.P., Hawley J.F., Krolik J.H., Hirose S.,
2005, ApJ, 620, 878

\reference{} Fabian A.C., Vaughan S., 2003, MNRAS, 340, L28

\reference{} Fabian A.C., Miniutti G., 2005, astroph/0507409

\reference{} Fabian A.C. et al., 1989, MNRAS, 238, 729

\reference{} Fabian A.C., Iwasawa K., Reynolds C.S., Young A.J., 2000,
PASJ, 112, 1145

\reference{} Fukumura K., Kazanas D., 2007, ApJ, in press
(astro=ph/0704.2159)

\reference{} Gammie C.F., 1999, ApJ, 522, L57

\reference{} King A.R., Kolb U., 1999, MNRAS, 305, 654

\reference{} Krolik J.H., 1999, ApJ, 515, L73

\reference{} Krolik J.H., Hawley J.F., 2002, ApJ, 573, 754

\reference{} Hawley J.F., 2000, ApJ, 528, 462

\reference{} Hawley J.F. Krolik J.H, 2001, ApJ, 548, 348

\reference{} Hawley J.F., Gammie C.F., Balbus S.A., 1996, ApJ, 464, 690

\reference{} Laor A.,  1991, ApJ, 376, 90

\reference{} McClintock J.E., Remillard R.A., 2003, astroph/0306213

\reference{} McClintock J.E., Shafee R., Narayan R., Remillard R.A.,
David S.W., Li L.-X., 2006, ApJ, 652, 518

\reference{} McKinney J.C., Gammie C.F., 2004, ApJ, 611, 977

\reference{} Merloni A., Vietri M., Stella L., Bini D., 1999, MNRAS, 304,
155

\reference{} Middleton M., Done C., Gierlinski M., Davis S.W., 2006,
MNRAS, 373, 1004

\reference{} Miller K.A., Stone J.M., 2000, ApJ, 534, 398

\reference{} Miniutti G., Fabian A.C., 2004, MNRAS, 349, 1435

\reference{} Miniutti G. et al., 2007, PASJ, 59, 315

\reference{} Moderski R., Sikora M., 1996, MNRAS, 283, 854

\reference{} Nowak M.A., Wagoner R.V., 1992, ApJ, 393, 697

\reference{} Nowak M.A., Wagoner R.V., Begelman M.C., Lehr D., 1997,
ApJL, 477, L91

\reference{} Novikov, I., Thorne K.S., 1973, in Black Holes, 1973,
ed. C.DeWitt and B.DeWitt(New York: Gordon and Breach).

\reference{} Paczynski B., Wiita P.J., 1980, A\&A, 88, 23

\reference{} Penrose R., 1969, Riv. de Nuovo Cim., 1, 252

\reference{} Pringle J.E., Rees M.J., 1972, A\&A, 21, 1

\reference{} Reynolds C.S., Armitage P.J., 2001, ApJL, 561, L81

\reference{} Reynolds C.S., Begelman M.C., 1997, ApJ, 488, 109 (RB97)

\reference{} Reynolds C.S., Ward M.J., Fabian A.C., Celotti A., 1997,
MNRAS, 291, 403

\reference{} Reynolds C.S., Nowak M.A., 2003, Phys. Reports, 377, 389

\reference{} Ross R.R., Fabian A.C., 2005, MNRAS, 358, 211

\reference{} Shafee R., Narayan R., McClintock J.E., 2007, ApJ, submitted
(astro-ph/0705.2244)

\reference{} Shafee R., McClintock J.E., Narayan R., Davis S.W., Li
L.-X., Remillard R.A., 2006, ApJ, 636, L133

\reference{} Shakura, N.I., Sunyaev, R.A., 1973, A\&A, 24,337

\reference{} Speith R., Riffert H., Ruder H, 1995,
Comput. Phys. Commun., 88, 109

\reference{} Strohmayer T.E., 2001, ApJ, 552, L49

\reference{} Stone J.M., Norman M.L., 1992a, ApJS, 80, 753

\reference{} Stone J.M., Norman M.L., 1992b, ApJS, 80, 791

\reference{} Turner N.J., 2004, ApJ, 605, L45

\reference{} Volonteri M. et al., 2005, ApJ, 620, 69

\reference{} Young A.J., Ross R.R., Fabian A.C., 1998, MNRAS, 300, L11
\end{references}
\end{document}